\documentclass[aps,prb, twocolumn ,showpacs,preprintnumbers, superscriptaddress]{revtex4}

\usepackage{graphicx}
\begin{document}
\preprint{}

\title{ Resistive relaxation in field-induced insulator-metal transition of a (La$_{0.4}$Pr$_{0.6}$)$_{1.2}$Sr$_{1.8}$Mn$_{2}$O$_{7}$ bilayer manganite single crystal}

\author{M.Matsukawa} 
\email{matsukawa@iwate-u.ac.jp }
\author{K.Akasaka}
\author{H.Noto}
\affiliation{Department of Materials Science and Technology, Iwate University , Morioka 020-8551 , Japan }
\author{R.Suryanarayanan}
\affiliation{Laboratoire de Physico-Chimie de L'Etat Solide,CNRS,UMR8648
 Universite Paris-Sud, 91405 Orsay,France}
\author{S.Nimori}
\affiliation{National Institute for Materials Science, Tsukuba 305-0047 ,Japan}
\author{M.Apostu}
\author{A.Revcolevschi}
\affiliation{Laboratoire de Physico-Chimie de L'Etat Solide,CNRS,UMR8648
 Universite Paris-Sud, 91405 Orsay,France}

\author{ N. Kobayashi }
\affiliation{Institute for Materials Research, Tohoku University, Sendai 980-8577, Japan }
\date{\today}

\begin{abstract}
We have investigated the resistive relaxation of a (La$_{0.4}$Pr$_{0.6}$)$_{1.2}$Sr$_{1.8}$Mn$_{2}$O$_{7}$ single crystal, in order to examine the slow dynamics of the field-induced insulator to metal transition of bilayered manganites.
The temporal profiles observed in remanent resistance follow  a  stretched  exponential  function accompanied by a slow relaxation similar to that obtained in  magnetization and magnetostriction data. We demonstrate that the remanent relaxation in magnetotransport  has a close relationship with magnetic relaxation that can be understood in the framework of an effective medium approximation by assuming that the first order parameter is proportional to the second order one.
\end{abstract}
\pacs{75.47.Lx,75.50.Lk}
\maketitle
\section{INTRODUCTION}

The discovery of the colossal magnetoresistance (CMR)  effect in doped manganites with perovskite structure has stimulated considerable interest for the understanding of their physical properties \cite{TO00}. Though the insulator to metal (IM) transition and its associated CMR are well explained  on the basis of  the double exchange (DE) model,  it is pointed out that the dynamic Jahn-Teller (JT) effect due to the strong electron-phonon interaction, plays a significant role in 
the appearance of CMR as well as the DE interaction \cite{ZE51,MI95}.  Furthermore, Dagotto $et\ al.$ propose a phase separation model where ferromagnetic (FM) 
metallic and antiferromagnetic (AFM) insulating clusters coexist as  supported by  recent experimental studies on the physics of manganites  \cite{DA01}. 
 
The bilayer manganite La$_{1.2}$Sr$_{1.8}$Mn$_{2}$O$_{7}$ exhibits a paramagnetic insulator (PMI) to  ferromagnetic metal (FMM) transition around  $T_{c}\sim$120K  and  its  associated CMR effect \cite{MO96}. In comparison with cubic manganites, the MR effect of the compound under consideration, due to its layered structure, is enhanced by  two orders of magnitude, at 8T, around $T_{c}$. It is well known that Pr-substitution on the La-site leading to (La$_{1-z}$Pr$_{z}$)$_{1.2}$Sr$_{1.8}$Mn$_{2}$O$_{7}$ causes an elongation of the $c$ axis length in contrast with  a shrinkage of the $a$($b$) axis, resulting in a change of the e$_{g}$-electron occupation from the d$_{x^2-y^2}$ to the d$_{3z^2-r^2}$ orbital \cite{MO97,OG00,WA03}. These findings also accompany a variation of the easy axis of magnetization from the $ab$ plane to the $c$ axis. For the $z$=0.6 crystal, the field-induced FMM state is realized, instead of the PMI ground state in the absence of magnetic field.   
In Fig.1, a phase diagram in  the  $(M,T)$ plane established from the magnetization measurements carried out on the $z=0.6$ crystal, with three regions labeled as the PMI,\ FMM and mixed phases (hatched area) is presented \cite{APDC}.  
A schematic diagram of free energy with two local minima corresponding to the PMI and FMM states is also given in Fig.1,  for  the virgin state (a) before application of the  magnetic field,  the  field-induced state (b) after the PMI to FMM transition and the mixed state (c)  after removal of the field.  Just after removing the field, the system still remains in a metastable FMM state. After a long time, the system comes back to the original PMI state through the mixed state consisting of both FMM and PMI regions.  In the mixed state, the total system is divided into a large number of subsystems which are randomly distributed with different local densities of free energy , causing complex relaxation processes observed  in the physical property studies \cite{AN99,UE00,LO01,DH01,GO01,MA04}. A magnetic frustration between  double-exchange ferromagnetic and superexchange antiferromagnetic  interactions at the Mn sites gives rise to a spin-glass-like behavior in manganites \cite{UE00,LO01,GO01}. In the mixed phase composed  of metallic and insulating regions, it is believed that the resistive relaxations reported \cite{AN99,DH01} arise from an electronic competition between double-exchange like itinerancy  and carrier localization associated with the formation of polarons.  
Recently, the slow dynamics of a remanent lattice striction of (La$_{0.4}$Pr$_{0.6}$)$_{1.2}$Sr$_{1.8}$Mn$_{2}$O$_{7}$ single crystal has been examined on the basis of a competition between Jahn-Teller type orbital-lattice and DE interactions\cite{MA04}. The former interaction induces a local lattice distortion of Mn O$_{6}$ octahedra along the $c$-axis but the latter suppresses a lattice deformation through the itinerant state \cite{ME99}.  Thus, it is desirable to establish a close relationship among the resistive, magnetic and lattice relaxations, for our understanding of  the CMR phenomena in bilayered manganites.  

\begin{figure}[ht]
\includegraphics[width=8cm]{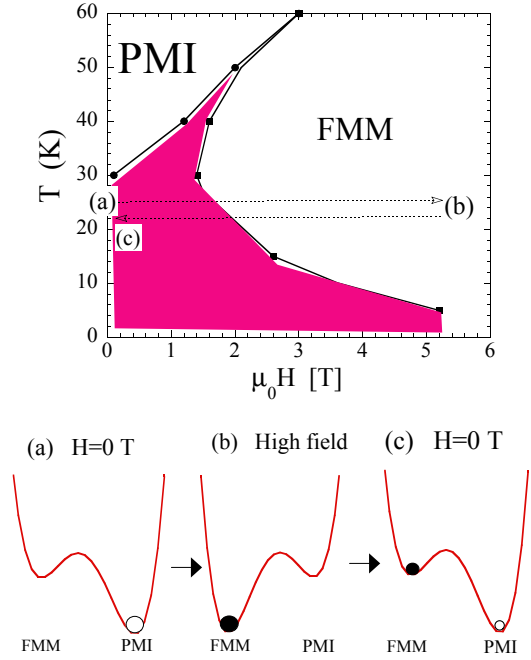}
\caption{Magnetic phase diagram in the ($H,T$) plane established from the magnetic measurements carried out on the $z$=0.6 crystal. A schematic picture of the free energy with two local minima corresponding to the FMM and PMI phases(regions (a),(b) and (c) correspond to   virgin state,\ field-induced metallic state  and\ mixed state, respectively.) }
\end{figure}%

Hence,  we have investigated the resistive relaxation of a (La$_{0.4}$Pr$_{0.6}$)$_{1.2}$Sr$_{1.8}$Mn$_{2}$O$_{7}$ single crystal. We compare our results with both magnetic and lattice relaxation data on  the $z$=0.6 crystal.
\section{EXPERIMENT}

Single crystals of (La$_{0.4}$Pr$_{0.6}$)$_{1.2}$Sr$_{1.8}$Mn$_{2}$O$_{7}$ were grown by the floating zone method using  a mirror furnace. 
The calculated lattice parameters were shown in a previous report \cite{AP01}.  The dimensions of  the $z$=0.6 sample are 3.4$\times$3 mm$^2$ in the $ab$-plane and 1mm along the $c$-axis.  Magnetoresistance was measured by means of a conventional four-probe technique at the Tsukuba Magnet Laboratory, the National Institute for Materials Science and at the High Field Laboratory for Superconducting Materials, Institute for Materials Research, Tohoku University. Magnetostriction measurements were performed using a strain gauge method\cite{MA04}. 
The magnetization measurements were made  using a superconducting quantum interference device  magnetometer at Iwate University.

\begin{figure}[ht]
\includegraphics[width=10cm]{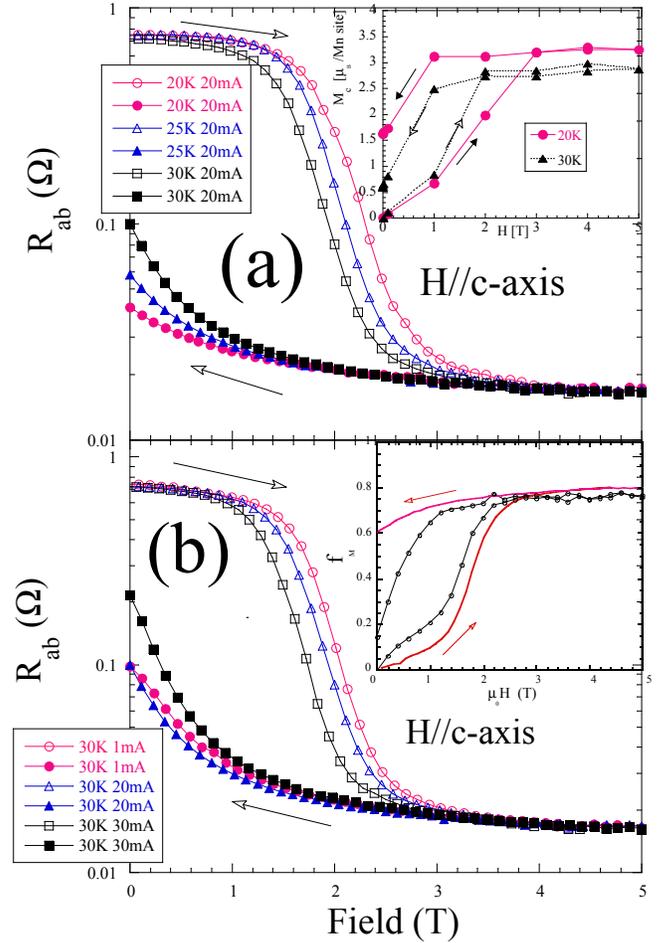}
\caption{ Magnetoresistance data $R_{ab}$ of a (La$_{0.4}$ Pr$_{0.6}$)$_{1.2}$Sr$_{1.8}$Mn$_{2}$O$_{7}$  single crystal in a field applied along the $c$ axis,  (a)at $T$ =\ 20, 25 and 30 K for $I$ = 20 mA  and (b) at $I$ = 1, 20 and 30 mA for 30 K. The inset of (a) shows the field dependence of the magnetization along the $c$ axis at  both 20 and 30 K .
In the inset of (b), a solid curve represents the volume fraction of metal phase $f_{M}$ estimated from the $R(H)$ data using an effective medium approximation discussed in the text. For comparison, the normalized magnetization curve $M(H)$ /$M_{full}$ at 30K is also presented. }
\end{figure}%

\section{RESULTS AND DISCUSSION}

 Let us show in Fig.2 the magnetoresistance data $R_{ab}$ of (La$_{0.4}$, Pr$_{0.6}$)$_{1.2}$Sr$_{1.8}$Mn$_{2}$O$_{7}$ single crystal at selected temperatures. Firstly, a field-induced insulator to metal transition and its associated CMR effect are observed around 2T, accompanied by a huge decrease in resistance by about two-orders of magnitude.  Secondly, a clear hysteresis in $R_{ab}$ is seen even though applied fields are lowered down to zero. As mentioned above, the system still remains in a metastable state just after the external field is switched off.  In Fig.2 (a), it can be seen that the characteristic field which switches the sample state from PMI to FMM, depends upon temperature and 
increases from 1.8T at 30K to 2.2T at 20K, in a good agreement with the magnetization curves in the inset of Fig.2 (a). Moreover, such a critical field is also controlled by changing an applied current (Fig.2 (b)) A local joule heating  assists a jump over potential barriers of local free energy allowing them to shift 
 from PMI towards FMM states at numerous PMI clusters within the sample, resulting in a suppression in both the switching field and hysteresis effect.  
Here, we estimate a volume fraction of metal (or insulator) from the $R(H)$ data using an effective medium approximation (EMA) \cite{BR35,BE93}.   
In our calculation, we assume a two-component composite material made up of both metallic and insulating grains with their resistivities, $\rho_{M}$ and $\rho_{I}$, giving an effective resistivity $\rho_{e}$ for spherical shape as follows;
\begin{equation}
f_{M}\frac{\rho_{e}-\rho_{M}}{\rho_{e}+2\rho_{M}}+ (1-f_{M})\frac{\rho_{e}-\rho_{I}}{\rho_{e}+2\rho_{I}}=0
\end{equation}

, where $f_{M}$ denotes a volume fraction of metal with  metallic resistivity $\rho_{M}$.  
Substituting the $R(H)$ data into the above equation and solving it with respect to $f_{M}$, we get a volume fraction of metal as shown in the inset of  Fig.2(b). For comparison, the magnetization curve at 30K  is also presented. The calculated curve based on the EMA  roughly reproduces the $M(H)$ curve except for the low-field region in the demagnetization process.  The difference in $M(H)$ and $f_{M}$ is probably related to the formation of magnetic domains conserving  ferromagnetic moments \cite{TO05}. 
Here,  $\rho_{I}$ is taken as the value of $R$ just before application of the field and  $\rho_{M}$ is determined from the value of $R(H)$ at a maximum field of 5T.  Furthermore, a volume fraction of metallic cluster at 5T is assumed to be equal to the ration $M$(5T)/$M_{full}$= $\sim$0.8 , in which $M_{full}$ means the value of full magnetization corresponding to the magnetic moment of the Mn ion(=3.4 $\mu_{B}$\ at a hole content $x$=0.4).  According to a previous work on cubic manganites by Jaime et al. \cite {JA99}, assuming both ferromagnetic and electronic free energy functionals and minimizing the total free energy, they obtain one solution where the first order parameter $m$ (=$M(H,T)$ /$M_{full}$) is proportional to the second order parameter $c$ (=$f_{M}$).  Thus, it is reasonable to take the preceding assumption in the EMA. 

 \begin{figure}[ht]
\includegraphics[width=8cm]{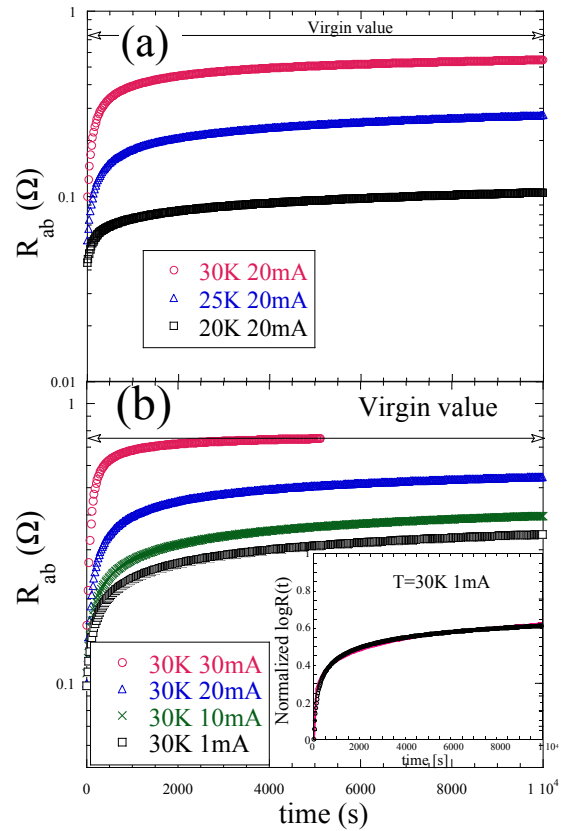}%
\caption{ Resistive relaxation profiles of  a (La$_{0.4}$Pr$_{0.6}$)$_{1.2}$Sr$_{1.8}$Mn$_{2}$O$_{7}$  single crystal as a function of  (a)temperature and (b) current . (a) $I$= 1,10,20 and 30mA at 30K,  (b) $T$=20,25 and 30K at $I$=30mA. The inset of (b) represents a typical curve fitted to normalized $R(t)$ data at 30K with $I$=1mA, using a stretched exponential function with the characteristic relaxation time and exponent, $\tau$ and $\beta$,respectively. We have $\tau$ =1.1$\times$10$^{4}$ s and $\beta$=0.25.
}
\end{figure}%

Now, we examine the resistive relaxation data as a function of temperature and excited current as depicted in Fig.3 (a) and (b). The system starts from a metastable state of the coexistence between metallic and insulating regions when a field is turned off , and should come back to a stable insulator at the original ground-state after a very long time . At 30K, the value of $R_{ab}$ with $I$=30 mA  rapidly relaxes within a few hundred seconds and then restores the ground-state value, as shown in Fig.3.  The relaxation time of  remanent $R_{ab}$ is elongated  at least by two orders of magnitude upon decreasing temperature from 30 K down to 20K.  We have noted from previous studies  that a relaxation curve in both  remanent magnetization and lattice striction in a (La$_{0.4}$Pr$_{0.6}$)$_{1.2}$Sr$_{1.8}$Mn$_{2}$O$_{7}$ single crystal is well fitted using a stretched exponential function with the characteristic relaxation time and exponent, $\tau$ and $\beta$.  A deviation in exponent from $\beta$=1 indicates the existence of multiple relaxation processes in the observed slow dynamics. 
In a similar way, we try to examine a temporal profile of remanent magnetoresistance following a stretched exponential form such as normalized log$R(t) = [$log$R(t)-$log$R(0)]/[$log$R(\infty)-$log$R(0)]$=1-exp$[-(t/\tau)^\beta]$, where $R(\infty)$ and $R(0)$ denote the virgin and initial values, before application of the field and just after removal of the field, respectively. A typical curve fitted to normalized $R(t)$ data at 30K with $I$=1mA is presented in the inset of Fig.3(b). 

\begin{figure}[ht]
\includegraphics[width=10cm]{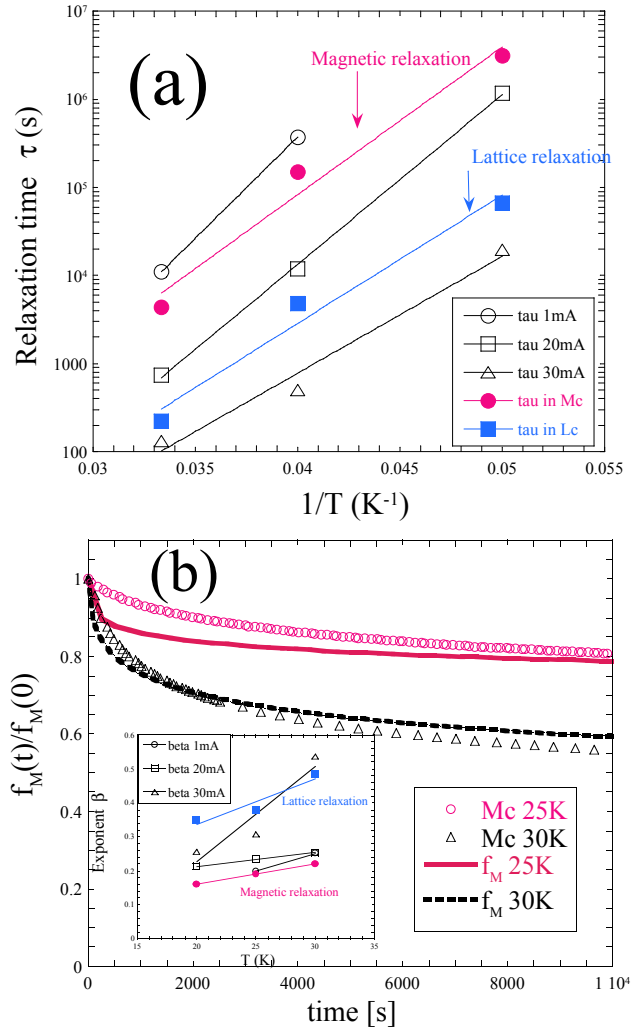}%
\caption{ (a) The resistive relaxation time $\tau_{R}$  as a function of $1/T$ for $I$= 1,20 and 30 mA. For comparison, both the magnetic and lattice relaxation parameters, $\tau_{M}$ and $\tau_{L}$, are also given.(b) A relaxation profile of the metallic fraction $f_{M}$ estimated from the $R(t)$ data using the EMA model. Solid and dashed curves represent calculation data at 25 and 30 K, respectively.  The normalized $c$ axis magnetization data, $M_{c}(t)/ M_{c}(0)$, are given. In the inset of (b), the exponent\ $\beta$ in the resistive, magnetic and lattice relaxations is plotted as a function of temperature. }
\end{figure}%

As a result, the fitted parameters $\tau_{R}$ and $\beta$ are plotted as a function of temperature, as shown in Fig. 4.  For comparison, the previous relaxation parameters, $\tau_{M}$ and $\tau_{L}$,   for both magnetization and magnetostriction curves are also given.  Firstly, upon decreasing the applied current,  the value of $\tau_{R}$  tends to approach the magnetic relaxation time , $\tau_{M}$. This tendency is also observed in the temperature variation of exponent\ $\beta$, as shown in the inset of Fig.4 (b).   On the other hand, the value of $\tau_{L}$ is smaller by about two orders of magnitude  than the lifetime of $R(t)$ and $M(t)$. 
Secondly, the relaxation time in $R(t), M(t)$ and $L(t)$ follows the thermally activated $T$-dependence, $\tau=\tau_{0}$exp($\Delta /kT$), where $\Delta$ denotes the activation energy corresponding to the  potential barrier between the metastable FMM state and the local maximum in free energy.  $\tau_{0}$ represents the intrinsic relaxation time determined from  microscopic mechanism. 
The activation energy of $R(t)$, $\Delta_{R}$ varies from 305 K at $I$=30mA, through 443K at $I$= 20mA, up to 530 K at $I$=1mA.  These values are not far from the activation energies of both remanent lattice and magnetization, $\Delta_{L}$=335 K and $\Delta_{M}$=386 K. The resistive and magnetic relaxations are  taken as signatures of the phase transition\ from metastable FMM to stable PMI states in the long time scale. On the other hand, the lattice relaxation is not due to the structural transition associated with cooperative phenomena but arises from a local lattice distortion of MnO$_{6}$ octahedra without a long-range order.   
In Fig.4 (b), a temporal profile of the metallic fraction $f_{M}$ estimated from the $R(t)$ data for  $I$=1mA using the  EMA model is given.  We notice that calculated curves of the metallic fraction  tend to approach the magnetization data after a long period of time.   This finding seems to be reasonable if we assume that  a ferromagnetic order parameter, $m$, is proportional to an electronic one, $f_{M}$. The difference in the initial drop between the metallic fraction and the magnetization curves is probably related to the formation of FMM domains might responsible for the disagreement observed  between $f_{M}$ and the normalized magnetization as depicted in the inset of Fig.2(b). 

\begin{figure}[ht]
\includegraphics[width=10cm]{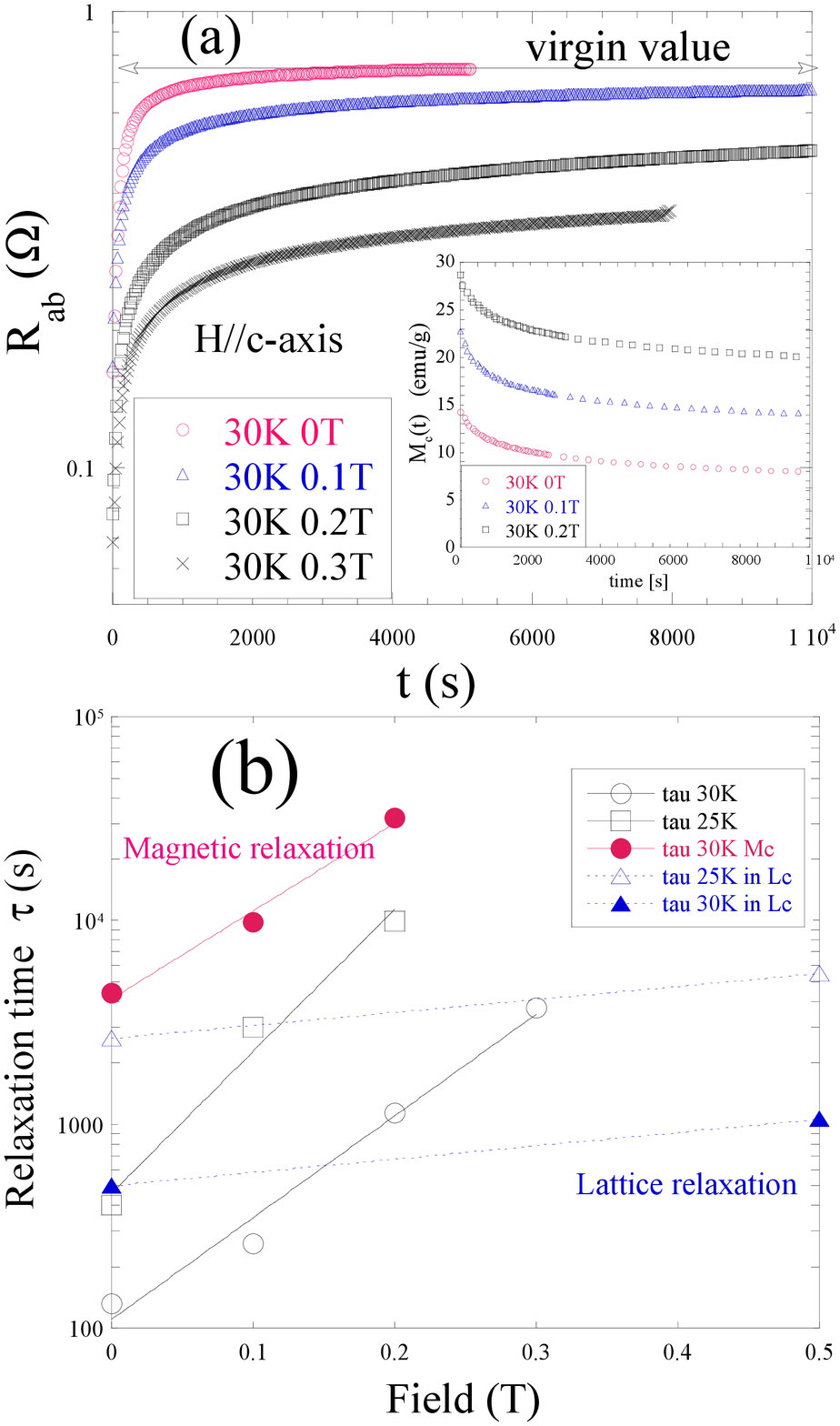}%
\caption{(a) Resistive relaxation profiles of  a (La$_{0.4}$Pr$_{0.6}$)$_{1.2}$Sr$_{1.8}$Mn$_{2}$O$_{7}$  single crystal as a function of field. $H$=0,0.1,0.2 and 0.3 T at 30K with $I$=30mA. In the inset of (a), magnetic relaxation data are also shown at 30K. 
(b) The resistive relaxation time $\tau_{R}(H)$  as a function of field at 25 and 30K with $I$= 30 mA. For comparison, the magnetic and lattice relaxation times, $\tau_{M}(H)$ and $\tau_{L}(H)$, are also given at selected temperatures.}
\end{figure}%

Finally, we explore resistive relaxation data as a function of field at selected temperatures, as shown in Fig.5.   The value of the relaxation time grows exponentially upon increasing the applied field because a local minimum in the free energy of the metastable state is stabilized by lowering the minimum free energy by $\mu_{eff}\mu_{0}H$. The field dependence of $\tau(H)$,  is well fitted by such a functional form as $\tau_{R}(0)$exp($\mu_{eff}\mu_{0}H$/$kT$). The effective magnetic moment $\mu_{eff}$ is expressed  as $\mu_{eff}=Ng\mu_{B}S$, giving  the average number of the Mn ions,\ $N$, contributing to the relaxation process of  the FMM to PMI transition at the level of clusters in divided subsystems \cite{AN99}.   Here,  $S$ represents the average spin number at the Mn ion site and we set  $S$ = 1.8 at a hole concentration of  0.4. A characteristic size of FMM clusters is estimated from the relaxation data using the exponential functional form to be $N_{R}$=140 at 30 K ($N_{R}$=166 at 25K). Moreover, from magnetic $\tau(H)$ we get $N_{M}$=122 at 30 K similar to the value of  $N_{R}$ . If the average distance between adjacent Mn ions is taken as 4 \AA, the cluster size of the FMM region reaches several tens of nanometer.  On the other hand, $\tau_{L}$ is independent of field up to 0.5T and shows no outstanding variation,\ in contrast with the value of both $\tau_{R}(H)$ and $\tau_{M}(H)$. 
This finding indicates that magnetostriction phenomena  are  not always associated with a long-rang order parameter although magnetization and magnetotransport are closely related to it.


In summary, we have shown that the field-induced insulator to metal transition observed in the single crystal of  (La$_{0.4}$Pr$_{0.6}$)$_{1.2}$Sr$_{1.8}$Mn$_{2}$O$_{7}$ is accompanied by a resistive relaxation process.
The temporal profiles observed in remanent resistance follow  a  stretched  exponential  function accompanied by a slow relaxation similar to those exhibited by  magnetization and magnetostriction. 
We demonstrate that the remanent relaxation in magnetotransport  has a close relationship with the magnetic relaxation that can be understood in the framework of an effective medium approximation assuming a proportionality between the  first  order parameter and the second order one.   

\begin{acknowledgments}
This work was partially supported by a Grant-in-Aid for Scientific Research from the Ministry of Education, Science and Culture, Japan. The authors thank  Dr.H. Ogasawara for his technical support. 
\end{acknowledgments}


\begin{thebibliography}{30}
\bibitem{TO00} $Colossal Magnetoresistive Oxides$, edited by Y.Tokura (Gordon and Breach,New York,2000).
\bibitem {ZE51}C.Zener,Phys.Rev.82,403(1951); P.G.deGennes,$ibid$.118,141 (1960).
\bibitem {MI95} A.J.Mills,P.B.Littlewood, and B.I.Shraiman,
 Phys.Rev.Lett.74,5144(1995); A.J.Mills, B.I.Shraiman, and R.Mueller, $ibid$.77,175 (1996).
\bibitem {DA01} For a recent review, see E.Dagotto, T.Hotta, and A.Moreo, 
Phys.Rep.344,1 (2001).
\bibitem {MO96}Y.Moritomo, A.Asamitsu, H.Kuwahara, and Y.Tokura,
 Nature 380,141 (1996).
\bibitem {MO97}Y.Moritomo, Y.Maruyama,T.Akimoto, and A.Nakamura
, Phys.Rev.B56,R7057(1997).
\bibitem{OG00}H.Ogasawara,M.Matsukawa,S.Hatakeyama,M.Yoshizawa,
M.Apostu, R.Suryanarayanan, G.Dhalenne, A.Revcolevschi, K.Ithoh, and N.Kobayashi, 
J.Phys.Soc.Jpn.69,1274(2000).
\bibitem {WA03}F.Wang,A.Gukasov,F.Moussa,M.Hennion, M.Apostu,
R.Suryanarayanan, and A.Revcolevschi, Phys.Rev.Lett.91,047204(2003).
\bibitem {AP01}M.Apostu, R.Suryanarayanan, A.Revcolevschi, 
H.Ogasawara, M.Matsukawa, M.Yoshizawa, 
and N.Kobayashi,Phys.Rev.B64,012407(2001).
\bibitem {APDC}M.Apostu, Doctoral thesis, Univ. of Paris-Sud(2002).
\bibitem {AN99}A.Anane,J.P.Renard,L.Reversat,C.Dupas,P.Veillet,
M.Viret,L.Pinsard, and A.Revcolevschi, Phys.Rev.B59,77(1999).
\bibitem {UE00}M.Uehara and S.W.Cheong, Europhys. Lett. 52,674(2000).
\bibitem {LO01}J.Lopez,P.N.Lisboa-Filho,W.A.C.Passos,W.A.Ortiz,
 F.M.Aruajo-Moreira, O.F. de Lima, D. Schaniel,
and K. Ghosh,Phys.Rev.B63,224422(2001).
\bibitem {DH01}J.Dho,W.S.Kim, and N.H.Hu, Phys.Rev.B65,024404(2001).
\bibitem {GO01}I.Gordon,P.Wagner,V.V.Moshchalkov,Y.Bruynseraede,
M.Apostu, R.Suryanarayanan,and A.Revcolevschi,
Phys.Rev.B64,092408(2001).
\bibitem {MA04}M.Matsukawa,M.Chiba,K.Akasaka, R.Suryanarayanan, 
 M.Apostu, A.Revcolevschi,S.Nimori, and N.Kobayashi, 
Phys.Rev.B70,132402(2004).
\bibitem {ME99}M.Medarde,J.F.Mitchell,J.E.Millburn, S.Short, 
and J.D.Jorgensen, Phys.Rev.Lett.83,1223(1999).
\bibitem {BR35}D.A.G.Bruggeman,Ann.Physik 24(1935)636.
\bibitem {BE93}D.J.Bergmann and D.Stroud, Solid State Phys.46 (1993) 147.
\bibitem {TO05}M.Tokunaga,Y.Tokunaga, and T.Tamegai, 
Phys.Rev.B71,012408(2005).
\bibitem {JA99}M.Jaime,P.Lin,S.H.Chun,M.B.Salamon,
P.Dorsey,and M.Rubinstein, Phys.Rev.B60,1028(1999).
\end{thebibliography}

\end{document}